\documentclass[journal]{IEEEtran}
\usepackage{cite}
\usepackage{url}
\usepackage{empheq}
\usepackage{float}
\usepackage{dsfont}

\usepackage{algorithm}
\usepackage{algorithmic}

\usepackage{amsmath,amssymb,amsfonts}
\allowdisplaybreaks[4]
\usepackage{graphicx}
\usepackage{textcomp}
\usepackage{amsmath}
\usepackage{enumerate}
\usepackage{epstopdf}
\usepackage{array}
\usepackage{booktabs}
\usepackage{subfigure}
\usepackage{multirow}
\usepackage{soul, color}
\usepackage[usenames,dvipsnames]{xcolor}
\usepackage[latin1]{inputenc}

\newtheorem{proposition}{Proposition}

\soulregister\cite7 
\soulregister\citep7 
\soulregister\citet7 
\soulregister\ref7 
\soulregister\pageref7 

\begin{document}

\title{Distributed Coordination of Charging Stations Considering Aggregate EV Power Flexibility}

\author{
Dongxiang~Yan,
Chengbin~Ma,~\IEEEmembership{Senior Member,~IEEE},
and Yue~Chen,~\IEEEmembership{Member,~IEEE}

\thanks{D. Yan and Y. Chen are with the Department of Mechanical and Automation Engineering, the Chinese University of Hong Kong, Hong Kong SAR, China (e-mail: dongxiangyan@cuhk.edu.hk, yuechen@mae.cuhk.edu.hk).

Chengbin Ma is with the University of Michigan-Shanghai Jiao Tong University Joint Institute, Shanghai Jiao Tong University, Shanghai 200240, China (e-mail: chbma@sjtu.edu.cn).
}}
\markboth{Journal of \LaTeX\ Class Files,~Vol.~XX, No.~X, Feb.~2019}%
{Shell \MakeLowercase{\textit{et al.}}: Bare Demo of IEEEtran.cls for IEEE Journals}

\maketitle

\begin{abstract}
In recent years, electric vehicle (EV) charging stations have witnessed a rapid growth. However, effective management of charging stations is challenging due to individual EV owners' privacy concerns, competing interests of different stations, and the coupling distribution network constraints. To cope with this challenge, this paper proposes a two-stage scheme. In the first stage, the aggregate EV power flexibility region is derived by solving an optimization problem. We prove that any trajectory within the obtained region corresponds to at least one feasible EV dispatch strategy. By submitting this flexibility region instead of the detailed EV data to the charging station operator, EV owners' privacy can be preserved and the computational burden can be reduced. In the second stage, a distributed coordination mechanism with a clear physical interpretation is developed with consideration of AC power flow constraints. We prove that the proposed mechanism is guaranteed to converge to the centralized optimum. Case studies validate the theoretical results. Comprehensive performance comparisons are carried out to demonstrate the advantages of the proposed scheme.
\end{abstract}


\IEEEpeerreviewmaketitle

\vspace{-0.cm}
\section{Introduction}
\IEEEPARstart {T}{he} proliferation of electric vehicles (EVs) has spurred the rapid development of EV charging stations \cite{wang2018electrical}. 
However, due to the random and relatively high EV charging power\cite{lee2021adaptive}, the sudden pulse-like and high charging load of the charging stations strains the distribution grid \cite{clement2009impact}.
Meanwhile, various techniques, such as intermittent renewable generation and distributed energy storage, begin to appear in the charging station and distribution network \cite{ref1}. This further complicates the operation of charging stations and threatens the power system reliability.
It is crucial to explore how to manage the growing number of charging stations and unlock the power flexibility hidden in EVs to jointly maintain system reliability.

Recently, there have been extensive literature on the optimal operation of multiple charging stations/microgrids to improve the system efficiency.
EV charging was scheduled to align with the local wind power generation of charging stations \cite{yang2018distributed}. But it did not account for the wind energy exchange between different stations, i.e., there was no energy trading between charging stations.
A rule-based strategy to operate multiple charging stations was proposed in \cite{yinhe}, where charging stations with higher battery storage state of charge (SOC) are enforced to deliver energy to others. But the wishes of charging station operators was neglected. At present, the EV charging station has to curtail surplus energy or sell it back to the grid at a low and location-independent feed-in tariff. This reduces the charging stations' earnings and the power losses caused by transactions are not taken into account.
A new market mechanism with properly designed prices is desired \cite{chen2022review}.
To model the interaction between participants, a leader-follower game based peer-to-peer (P2P) scheme was widely adopted for sharing of facilities \cite{3p} and among multiple PV prosumers \cite{2017sg}.
But it may lead to sub-optimal results because of the competitive relationship. 
Energy trading optimization schemes designed for interconnected microgrids were proposed in \cite{wang2018incentivizing,cui2019building}. To reduce the dependence on the grid, microgrids with surplus energy directly trade with others with energy deficiency.
The energy trading model was further extended to the field of multi-energy systems \cite{jing2020fair}.
However, the above studies mainly focus on the design of energy trading mechanism without consideration of distribution network constraints.

Several studies began to integrate the distribution network model into the energy trading to better match the reality.
A practical distribution network model was incorporated in the energy sharing problem of microgrids \cite{liu2015energy}.
A centralized energy trading framework of EV charging stations was proposed in \cite{Affolabi2022optimal} that integrated a power distribution network.
The P2P energy trading problem of microgrids was formulated as a bi-level programming, which was then transformed into a single level optimization that can be solved centrally \cite{nezamabadi2020arbitrage}.
The above centralized schemes may jeopardise the privacy of charging stations by gathering their private data to make a central decision \cite{liu2017distributed}.
Therefore, distributed operation is preferred.
To protect the privacy of microgrids, a distributed algorithm based on subgradient method was proposed in \cite{liu2018hybrid}.
Compared with other distributed algorithms, the alternating direction method of multipliers (ADMM) based approach with good convergence properties and scalability \cite{2011boyd} has attracted much attention.
A distributed mechanism based on ADMM was proposed to determine the amount of energy traded among networked charging stations \cite{zhang2021distributed} and interconnected microgrids \cite{2020fixLoad}.
Similarly, the energy sharing problem of prosumers in \cite{guo2021asynchronous} adopted the traditional ADMM method and considered a simplified LinDistFlow model.
The distributed optimization based methods can achieve social optimum, but why participants should follow the decision rules obtained from the Lagrangian function is not clear.
Nash bargaining based model~\cite{fan2018bargaining} and mid-market ratio mechanism~\cite{tushar2018peer} were also applied to allocate the payment among participants.
Though distribution network constraints have been considered, the amount of shared energy between a pair of charging stations was assumed to be equal. This ignores the power losses during transmission and the system operator has to compensate for power losses.

Another important issue related to charging station management is how to take full advantage of the flexibility of EVs while protecting the privacy of the vehicle owner. Ref. \cite{2019YanQ} solved a centralized optimization problem to determine EV charging schedules and to manage the power balance in the charging station.
A centralized scheme based on improved particle swarm optimization was presented in \cite{qi2018pso} to schedule EVs' charging for cost reduction.
However, the above methods require the charging station to know the data of all EVs, which may violate the privacy of EV owners. In addition, with an increasing number of EVs, solving an optimization problem with detailed EV data and constraints will be time consuming \cite{zhang2017optimal}. An alternative approach is to derive the aggregate flexibility of the large number of resources first, which can simplify further analysis \cite{chen2020aggregate}. 
The aggregate flexibility of thermostatically controllable loads was explored in \cite{zhao2017geometric} that aims at providing ancillary services to the grid. The heterogeneous distributed resources' aggregate flexibility were modeled in \cite{xu2016hierarchical}.
Despite the above efforts, how to characterize the maximal potential power flexibility of EVs in a charging station remains to be investigated.
In particular, EV charging stations has some constraints distinct from other resources, such as the limitation due to a finite number of chargers.


Inspired by the above discussions, in this paper, we propose a two-stage scheme for coordinating EV charging stations in a distribution network. The first stage focuses on the characterization of aggregate EV power flexibility inside each charging station, so that we can make use of EVs' flexibility while preserving EV owners' privacy. In the second stage, a distributed coordination mechanism that can achieve social optimum with consideration of transmission losses is proposed. Our main contributions are three-fold:
\begin{enumerate}
  \item We propose an optimization problem to characterize the aggregate power flexibility of all EVs inside a charging station. It generates the upper and lower power trajectories of the aggregate flexibility region. We prove that any trajectory within this region corresponds to at least one feasible EV dispatch strategy. Thus, when managing its energy, the charging station can adjust the aggregate power of EVs within this region without knowing the detailed information of each EV. This can protect individual EV owner's privacy. Moreover, the challenging computational complexity of managing a large number of EVs can be resolved. 
  \item We propose a novel distributed coordination mechanism that uses a price that accounts for power losses as a medium to reconcile the energy that the charging station wishes to trade with the distribution system operator's trading schedule. We prove that the proposed mechanism is guaranteed to converge to the centralized optimum.
  \item Compared with the traditional method, the proposed method can achieve significant total cost savings, reduce power losses, and greatly improve the utilization of battery storage in charging stations. 
\end{enumerate}

The remainder of this paper is organized as follows. Section \ref{sec:model} describes the overall system structure, challenges, and the proposed two-stage scheme. Section \ref{sec:stage1} and \ref{sec:stage2} introduce the aggregate EV power flexibility evaluation problem and the distributed coordination mechanism for charging stations, respectively.
Simulation results are presented in Section \ref{sec:result}. Finally, Section \ref{sec:conclu} concludes this paper.

\section{Overview of the System and Challenges}\label{sec:model}
\begin{figure}[!htbp]
  \centering
  \includegraphics[width=0.45\textwidth]{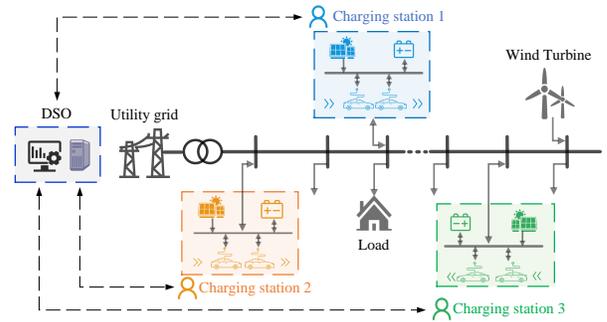}\\
  \caption{Structure of the EV charging stations system in a distribution network.}\label{fig:sysConf}
\end{figure}

Fig.~\ref{fig:sysConf} shows the overall structure of the studied multiple charging stations system.
All charging stations are equipped with PV panels and battery storage, but they may have different capacities and EVs' charging patterns.
We define $\mathcal{I}$ as the set of charging stations in the system, each of which is indexed by $i\in\mathcal{I}$.
The charging stations are located at different buses of a radial distribution network.
There is a distribution system operator (DSO) that monitors the power flow of the distribution network and the energy trading with the utility grid (connected to Bus 1).
Let $\mathcal{T}=\{1,...,T\}$ denote the operation time horizon and $T=24$.
Each time slot, indexed by $t\in\{1,...,T\}$, has an equal time interval $\Delta t$, i.e., 1 hour.

To ensure the reliable and efficient operation of the charging stations system, it needs to coordinate well the numerous energy resources linked by the charging stations. This is complicated in three ways: 1) \emph{Agents with conflicting interests}. For example, EV owners need to satisfy their charging needs, the charging station operator (CSO) aims to minimize its own operation cost, and the DSO has to ensure system security while minimizing power loss. 2) \emph{Privacy protection and information asymmetry.} The EV owners are not willing to reveal their private information (such as the charging needs) to the CSO; the distribution network constraints is only available to DSO; the DSO does not know the batteries and PV generations in the charging stations. 3) \emph{Computational complexity}. It is time-consuming for the CSO to manage the numerous resources inside it (the large number of EVs, PV, battery storage, etc.) as well as deciding on the energy trading plan by solving a centralized optimization problem.

To address the above challenges, we propose a two-stage energy management scheme. In the first stage, the aggregate power flexibility region (explain in detail later) of all EVs inside each charging station is derived. This allows the CSO to know the adjustable capability of EVs while protecting the privacy of individual EV owners. This aggregate EV power flexibility region serves as a constraint for the charging station's energy management in the second stage, which can also reduce computational burden. In the second stage, a distributed coordination mechanism is proposed to facilitate energy trading among charging stations and with the utility grid while ensuring the satisfaction of network constraints. The proposed mechanism is consistent with the asymmetric information structure that network constraint is only available to the DSO while the charging station's operational constraints known solely to the CSO. Each stage will be discussed in detail in the following sections.




\section{Aggregate EV Power Flexibility Evaluation-The 1st Stage}\label{sec:stage1}

In this section, we first introduce the concept of aggregate EV power flexibility and then propose an optimization problem to approximate it. The obtained approximated aggregate power flexibility region will be used in the second-stage charging station coordination problem, when the actual EV dispatch strategies will be determined.

\subsection{Aggregate EV Power Flexibility}
\begin{figure}[htbp]
  \centering
  \includegraphics[width=0.45\textwidth]{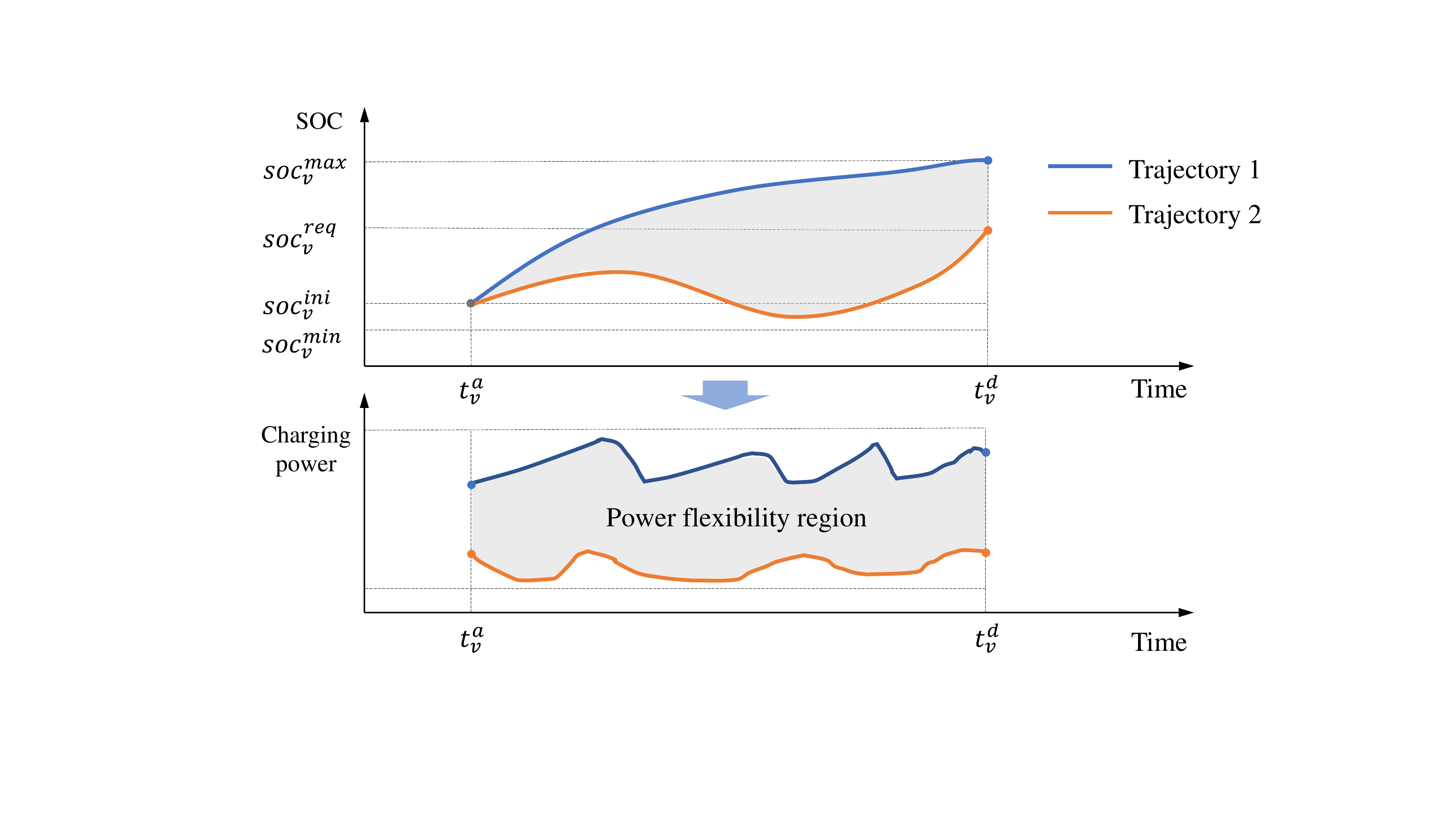}\\
  \caption{A diagram shows the power flexibility of an EV.}\label{fig:evublb}
\end{figure}

Let $\mathcal{V}_i$ denote the set of EVs that will arrive at and charge in the charging station $i\in\mathcal{I}$ in the next day.
For a certain EV $v\in\mathcal{V}_i$, its charging need can be defined by four parameters: $(t_v^a, t_v^d, soc_v^{ini}, soc_v^{req})$, where $t_v^a$ is its arrival time, $t_v^d$ is the anticipated departure time which should meet $t_v^a<t_v^d$, $soc_v^{ini}$ is its initial battery SOC level, and $soc_v^{req}$ represents the required least SOC when it leaves.
The EV will submit its charging need to the charging station one day before to make a reservation. Instead of a fixed load curve, the charging station operator has some flexibility in charging the EVs. For example, Fig.~\ref{fig:evublb} gives two possible ways to meet the charging need of an EV. Let $\{p_v^c(t),\forall t\}$ be the charging power of an EV $v$ over time, then the range that the charging power can vary within is called the \emph{EV's power flexibility}. If we sum up the power flexibility of all EVs in a charging station $i \in \mathcal{I}$, then we can get the \emph{aggregate EV power flexibility} of the charging station $i \in \mathcal{I}$, denoted by $\mathcal{F}_i$.

However, unlike the controllable generators, whose flexibility can be described by their minimum and maximum power at each time slot, it is hard to characterize the power flexibility of an EV since its SOC is temporally coupled. The power flexibility region in the current time slot can affect that in the next time slot.
To avoid no power flexibility being available during certain periods, we prefer a region with more evenly distributed power flexibility across all EVs' declared charging time slots, rather than having wide flexibility in some time slots but little or none in other time slots. Besides, we hope that any trajectory within the region corresponds to at least one feasible EV dispatch strategy that satisfies all the constraints related to charging in the station.


\subsection{Approximation of Aggregate EV Power Flexibility}
For ease of use, we aim to approximate the aggregate power flexibility of EVs in a charging station $i \in \mathcal{I}$ (denoted by $\mathcal{F}_i$) by a region $\tilde{\mathcal{F}}_i$ consisting of a series of intervals $[\check{p}_{d,i}(t),\hat{p}_{d,i}(t)]$ for each time slot $t \in \mathcal{T}$. That's $\mathcal{F}_i \approx \tilde{\mathcal{F}}_i:= [\check{p}_{d,i}(1),\hat{p}_{d,i}(1)] \times ... \times [\check{p}_{d,i}(T),\hat{p}_{d,i}(T)]$, so that the aggregate power flexibility can be specified by a lower power trajectory $\{\check p_{d,i}(t),\forall t\}$ and an upper power trajectory $\{\hat p_{d,i}(t),\forall t\}$. 
We formulate the following optimization problem to obtain the lower and upper power trajectories of the feasible region. We will prove later that for any $\{p_{d,i}(t),\forall t\} \in \tilde{\mathcal{F}}_i$, there always exists a feasible EV dispatch. 
\begin{subequations}
\label{eq:flexibility}
\begin{align}
\max \quad & \sum_{t \in \mathcal{T}} \left(\hat{p}_{d,i}(t)-\check{p}_{d,i}(t)-w(\hat{p}_{d,i}(t)-\check{p}_{d,i}(t))^2\right),\label{equ:drobj}\\
\hbox{s.t.}\quad
  &\hat{p}_{d,i}(t)=\sum_{v\in\mathcal{V}_i}\hat{p}^c_v(t),\forall t \label{equ:ubpdi}\\
  &-\hat{s}_v^c(t)p_{chg}^v\leq \hat{p}_v^{c}(t)\leq \hat{s}_v^c(t)p_{chg}^v,\forall v,\forall t\label{equ:ubpd}\\
  &0\leq\hat{s}_v^c(t)\leq s_v^p(t),\forall v,\forall t \label{equ:ubscsd}\\
  &\hat{soc}_v(t_v^a)= soc_v^{ini},\hat{soc}_v(t_v^d)\geq soc_v^{req},\forall v\label{equ:ubsntatd}\\
  &\hat{soc}_v(t+1)= \hat{soc}_v(t)+\frac{\hat{p}_v^{c}(t)\Delta t}{E_v},\forall v,\forall t \ne T\label{equ:ubsoc}\\
  &soc^{min}_v\leq \hat{soc}_v(t)\leq soc^{max}_v,\forall v,\forall t \label{equ:ubsocranges}\\
  &\sum_{n\in\mathcal{V}_i}\hat{s}_v^c(t)\leq N_{chg,i},\forall t\label{equ:ubnchg1}\\
  &\check{p}_{d,i}(t)=\sum_{n\in\mathcal{V}_i}\check{p}^c_v(t),\forall t\label{equ:lbpdi}\\
  &-\check{s}_v^c(t)p_{chg}^v\leq \check{p}_v^{d}(t)\leq \check{s}_v^c(t)p_{chg}^v,\forall v,\forall t\\
  &0\leq\check{s}_v^c(t)\leq s_v^p(t),\forall v,\forall t \label{eq:1}\\
  &\check{soc}_v(t_v^a)= soc_v^{ini},\check{soc}_v(t_v^d)\geq soc_v^{req},\forall v\\
  &\check{soc}_v(t+1)= \check{soc}_v(t)+\frac{\check{p}_v^{c}(t)\Delta t}{E_v},\forall v,\forall t \ne T\label{equ:lbsoc}\\
  &soc^{min}_v\leq \check{soc}_v(t)\leq soc^{max}_v,\forall v,\forall t\\
  &\sum_{n\in\mathcal{V}_i}\check{s}_v^c(t)\leq N_{chg,i},\forall t,\label{equ:lbnchg1} \\
  & \hat p_{d,i}(t) \ge \check p_{d,i}(t),\forall t; \hat s_v^c(t)=\check s_v^c(t),\forall v,\forall t. \label{equ:joint}
\end{align}
\end{subequations}
where
\begin{align}
   &s^p_v(t)=\left\{
  \begin{array}{ll}
  1 & \text{if }{t\in[t_v^a,t_v^d]}\\ 
  0 & \text{if }{t<t_v^a\cup t\geq t_v^d}
  \end{array} \right.,\forall v,\forall t.\label{equ:sp}
\end{align}

The objective function (\ref{equ:drobj}) is to maximize the aggregate power flexibility of EVs while the quadratic term makes the flexibility be distributed more evenly across time, and $w$ is a weight parameter.
Constraint (\ref{equ:ubpdi}) defines the upper bound of the aggregate EV power flexibility. The charging power of an EV $v$ is limited by (\ref{equ:ubpd}), where $p_{chg}^v$ is the maximum charging/discharging power. $s_{v}^c(t)$ indicates the charging status of an EV $v$ at time slot $t$. If the EV is being charged, $s_{v}^c(t)=1$; otherwise, $s_{v}^c(t)=0$. Constraints
(\ref{equ:ubscsd}) and (\ref{equ:sp}) ensure that charging only happens during the EV's declared parking time. If the EV is in the charging station, $s_{n}^p(t)=1$; otherwise, $s_{n}^p(t)=0$.
(\ref{equ:ubsntatd}) gives the EV's initial SOC and the charging requirement.
(\ref{equ:ubsoc}) and (\ref{equ:ubsocranges}) describe the EV's SOC dynamics and SOC range. $E_v$ is the EV battery capacity. Constraint
(\ref{equ:ubnchg1}) ensures that the number of EVs being charged at any time slot cannot exceed the number of chargers~$N_{chg,i}$ installed at the charging station $i$.
Similarly, (\ref{equ:lbpdi})-(\ref{equ:lbnchg1}) are the constraints related to the lower bound of the aggregate power flexibility. \eqref{equ:joint} is the joint constraint to ensure that any aggregate power trajectory between $\{\hat p_{d,i}(t),\forall t\}$ and $\{\check p_{d,i}(t),\forall t\}$ is achievable, which is formally stated in the proposition below.

\begin{proposition} \label{prop-1}
Let $\{\hat p_{d,i}^*, \check p_{d,i}^*,\forall t\}$ be the optimal solution of \eqref{eq:flexibility}. For any aggregate power trajectory $\{p_{d,i}(t),\forall t\}$ that satisfies $p_{d,i}(t) \in [\check p_{d,i}^*(t), \hat p_{d,i}^*(t)]$ for all time slots $t \in \mathcal{T}$, there exists a feasible EV dispatch strategy.
\end{proposition}

The proof of Proposition \ref{prop-1} can be found in Appendix \ref{appendix-A}. In practice, the aggregate EV power flexibility $\tilde{\mathcal{F}}_i$ can be generated by a smart EV management system using the above optimization problem \eqref{eq:flexibility} and submitted to the charging station operator for further use. The operator only knows the aggregate flexibility but not the detailed information of each EV, which can protect EV owner's privacy to some extent.


\section{Charging Station Coordination-The 2nd Stage}\label{sec:stage2}

With the aggregate EV power flexibility of each charging station $i \in \mathcal{I}$, in this section, we formulate the charging stations coordination problem, and propose a suitable distributed coordination mechanism to solve it.

\subsection{Modelling of Charging Station}
\subsubsection{EV Charging Demand}
The final dispatched aggregate charging demand in the charging station $i$ should be within the flexibility region $\tilde{\mathcal{F}}_i$ provided by the first stage:
\begin{equation}\label{equ:evcons}
    \check{p}_{d,i}(t)\leq p_{d,i}(t)\leq \hat{p}_{d,i}(t),\forall t.
\end{equation}
The charging power dispatched by the charging station may be less than the upper boundary, which may lead to dissatisfaction of EVs or decrease their utility \cite{2017sg}. We give a measure of this incurred dissatisfaction cost as
\begin{equation}
    C_{ev,i}(\mathbf{p_{d,i}})=\sum_{t\in\mathcal{T}}c_{cs,i}(\hat{p}_{d,i}(t)-p_{d,i}(t))
\end{equation}
where $c_{cs,i}$ is the cost coefficient.


\subsubsection{Battery Operation}
We let $p^{d}_{b,i}(t)$ and $p^{c}_{b,i}(t)$ denote the discharging and charging power of the battery energy storage in charging station $i\in\mathcal{I}$ at time slot~$t$.
The $p^{d}_{b,i}(t)$ and $p^{c}_{b,i}(t)$ should meet the following physical constraints:
\begin{align}
  &0\leq p^{d}_{b,i}(t)\leq p_{b,i}^{d,max},\label{equ:pbdch}\\
  &0\leq p^{c}_{b,i}(t)\leq p_{b,i}^{c,max},\label{equ:pbcha}
\end{align}
where $p_{b,i}^{d,max}$ and $p_{b,i}^{c,max}$ are the maximum discharging and charging power, respectively.

Along with battery discharging and charging, the battery storage SOC dynamics $soc_{b,i}$ can be expressed by:
\begin{equation}\label{equ:Ebat}
soc_{b,i}(t+1)=soc_{b,i}(t)-\frac{p^{d}_{b,i}(t)\Delta t}{\eta_dE_{b,i}}+\frac{p^{c}_{b,i}(t)\Delta t\eta_c}{E_{b,i}},
\end{equation}
where $E_{b,i}$ is the energy capacity of battery storage when it is fully charged; $\eta_d$/$\eta_c$ is the discharging/charging efficiency.
The SOC should always be within its allowable range to guarantee no over-discharging or over-charging occurs:
\begin{equation}\label{equ:EbatInequ}
soc^{min}_{b,i}\leq soc_{b,i}(t)\leq soc^{max}_{b,i},
\end{equation}
where $soc_{b,i}^{min}$ and $soc_{b,i}^{max}$ are the minimal and maximal SOC levels, respectively.
Besides, the battery SOC at the initial ($t=1$) and final ($t=T$) time slots are restricted to be equal so that the battery operation decouples across different days
\begin{equation}\label{equ:Ebatt}
soc_{b,i}(1)= soc_{b,i}(T).
\end{equation}

Both charging and discharging can cause battery degradation and the degradation cost is described by:
\begin{equation}\label{equ:batcost}
C_{b,i}(\mathbf{ p^d_{b,i}, p^c_{b,i}})=c_{b,i}\sum\limits_{t\in\mathcal{T}}\left(p^d_{b,i}(t)+p^c_{b,i}(t)\right)\Delta t,
\end{equation}
where $\mathbf{ p^d_{b,i}}=\{p^d_{b,i}(t),\forall t\in\mathcal{T}\}$, $\mathbf{ p^c_{b,i}}=\{p^c_{b,i}(t),\forall t\in\mathcal{T}\}$, $c_{b,i}$ is the degradation cost coefficient.

\subsection{Energy Trading with Other Charging Stations/Utility Grid}
In addition to using the local PV generation and battery storage, charging stations can also trade energy with other charging stations or the utility grid. In general, when the local PV generation exceeds the charging demand, the charging station can sell the surplus energy; otherwise, it can buy energy from other charging stations or the utility grid. These energy trading are performed on the distribution network.
\subsubsection{Conventional Method}
Conventionally, we let the charging station to buy/sell energy from/to the utility grid under a uniform price $\lambda^b_g(t)/\lambda^s_g(t)$.
The cost for the charging station $i\in\mathcal{I}$ to trade energy at time slot $t$ is calculated as follows:
\begin{equation}\label{equ:cgiBase}
C^{base}_{g,i}(\mathbf{p^b_{g,i},p^s_{g,i}}) =\sum_{t\in\mathcal{T}} \left(p^b_{g,i}(t)\lambda^b_g(t)-p^s_{g,i}(t)\lambda^s_g(t)\right)\Delta t,
\end{equation}
where $p^b_{g,i}(t)/p^s_{g,i}(t)$ is the power bought/sold from/to the utility grid.



However, since the charging stations are located at different buses and the power transmissions in distribution network result in power losses, giving them the same price is unfair. Also in that case, the DSO may need to pay for the power losses. To cope with these problems of the conventional method, we propose a new approach to better design the energy trading prices as follows.

\subsubsection{Proposed Method}
Due to the power loss of the network and the power flow limits, to be fair, the energy trading prices at different nodes should be different. To facilitate the energy trading, here we assume that the DSO will help to coordinate the trading by announcing an energy trading price $\lambda_{p,i}(t)$ to each charging station that takes into account multiple factors such as the grid buying/selling prices, the distance from the utility grid and other charging stations, etc. Upon receiving the price, the charging station will decide on how much energy it would like to trade. We let $p_{g,i}$ represent the net exchange power imported by the charging station. It is the sum of power bought from all other charging stations and from the utility grid. $p_{g,i}(t)$ is negative when the charging station sells.
The $p_{g,i}(t)$ should satisfy the following physical limits:
\begin{align}
  p_{g,i}^{\min}\leq p_{g,i}(t)\leq p_{g,i}^{\max},\label{equ:pgb}
\end{align}
where $p^{\min}_{g,i}$ and $p^{\max}_{g,i}$ are the minimum and maximum power imported by the charging station $i$. If $p_{g,i}(t)>0$, it means that the charging station $i\in\mathcal{I}$ buys energy, and vice versa.

Given the locational energy trading price $\lambda_{p,i}(t)$, the cost for the charging station $i\in\mathcal{I}$ to trade energy at time slot $t$ is calculated as follows:
\begin{equation}\label{equ:csgridcost}
C_{g,i}(\mathbf{ p_{g,i}})=\sum\limits_{t\in\mathcal{T}}\left(p_{g,i}(t)\lambda_{p,i}(t)\right)\Delta t,
\end{equation}
where the vector $\mathbf{ p_{g,i}}=\{p_{g,i}(t),\forall t\in\mathcal{T}\}$.

So far, we can formulate the total operation cost of charging station $i$, which consists of the above mentioned terms
\begin{equation}\label{equ:objci}
C_i(\mathbf{x_i}) = C_{b,i} +C_{ev,i}+C_{g,i},
\end{equation}
where vector $\mathbf{x_i}=\{\mathbf{p_{d,i},p_{g,i}, p^{d}_{b,i}, p^{c}_{b,i}}\}$ summarizes the decision variables of charging station $i$.

In addition, each charging station is assumed to be equipped with renewables such as PV panels for reducing the energy bought from the grid. The PV power generation at time $t$ is represented by $p_{pv,i}(t), i\in\mathcal{I}$. The charging station should maintain its internal power balance at each time slot $t$:
\begin{equation}\label{equ:pBala}
\begin{aligned}
p_{d,i}(t) = & p_{g,i}(t) + p_{pv,i}(t) + p^d_{b,i}(t) -p^c_{b,i}(t).
\end{aligned}
\end{equation}

\subsection{Modelling of the Distribution Network}

Considering that the charging stations are located at different buses of a radial distribution network, their energy trading cannot be achieved without the support of the power network. Thus, it is necessary to take into account the distribution network model.
Typically, the distribution network can be modeled as a graph $\mathcal{G(N,E)}$, where $\mathcal{N}$ and $\mathcal{E}$ are the set of buses and lines, respectively.
Then, we can index each bus in $\mathcal{N}$ by $n=1, 2, ... ,N$, and a branch can be represented by $(n,j)\in\mathcal{E}$.
A branch flow model is adopted \cite{low2013branch}:
\begin{subequations}
\label{eq:ACpowerflow}
\begin{align}
&p_j(t)=P_{nj}(t)-r_{nj}\ell_{nj}(t)-\sum_{k:(j,k)\in\mathcal{E}}P_{jk}(t),\label{equ:pinpf}\\
&q_j(t)=Q_{nj}(t)-x_{nj}\ell_{nj}(t)-\sum_{k:(j,k)\in\mathcal{E}}Q_{jk}(t),\\
&v_j(t)=v_{n}(t)-2(r_{nj}P_{nj}(t)+x_{nj}Q_{nj}(t))+(r^2_{nj}+x^2_{nj})\ell_{nj},\label{equ:pfvj}\\
&\ell_{nj}(t)=\frac{P_{nj}(t)^2+Q_{nj}(t)^2}{v_n(t)},\label{equ:pfnonconvex} \\
&\underline{p_j}\leq p_j(t)\leq \overline{p_j},\quad
\underline{q_j}\leq q_j(t)\leq \overline{q_j},\label{equ:pfpqub}\\
&0\leq \ell_{nj}(t)\leq \overline{\ell_{nj}},\quad
\underline{v_j}\leq v_j(t)\leq \overline{v_j},\label{equ:pflvub}
\end{align}
\end{subequations}
where $p_j$ and $q_j$ are the active and reactive power at bus $j\in\mathcal{N}$, $P_{nj}$ and $Q_{nj}$ are the active and reactive power flow in line $(n,j)$, $r_{nj}$ and $x_{nj}$ are the resistance and reactance in line $(n,j)$, $\ell_{nj}$ is the squared current $I_{nj}$ magnitude in line $(n,j)$, i.e., $\ell_{nj}=|I_{nj}|^2$, $v_{j}$ is the squared voltage $V_j$ magnitude at bus $j$, i.e., $v_j=|V_j|^2$, $\underline{\bullet}$ and $\overline{\bullet}$ represent the lower and upper bounds of the variable $\bullet$.

Let $n_i$ denote the bus that charging station $i$ is connected to. For the bus $n_i$, there is
\begin{align}
    p_{n_i}(t) = p_{g,i}(t),:\lambda_{p,i}(t)\label{equ:pi}
\end{align}
where $\lambda_{p,i}(t)$ is the dual variable of this equality, which is also the locational energy trading price. This equation builds the coupling connection between the charging station and the distribution network.

Bus 1, i.e., the slack bus in the distribution network, is responsible for buying/selling energy from/to the utility grid.
We denote $p_{1}^b(t)$/$p_{1}^s(t)$ as the energy bought/sold by bus 1 at time slot $t$.
Thus, the incurred cost is
\begin{equation}\label{equ:utilitycost}
C_{bus1}(t)=\left(p^b_{1}(t)\lambda^b_{g}(t)-p^s_{1}(t)\lambda^s_{g}(t)\right),
\end{equation}
where $\lambda^b_{g}(t)$ and $\lambda^s_{g}(t)$ are the utility electricity buy and sale prices, respectively.
Further, they should meet the requirement $\lambda^s_{g}(t)<\lambda^b_{g}(t)$ to ensure that the distribution network won't arbitrage by buying from and selling back to the utility at the same time.

Generally, the DSO is in charge of the distribution network management and will regulate the energy transactions among charging stations. Here, we use the minimization of utility grid cost plus power loss and minus the revenue of selling electricity to charging stations as the objective function of DSO,
\begin{equation}\label{equ:objdso}
\begin{aligned}
    \min\quad &C_{dso}(\mathbf{x_{d}})
    =C_{bus1}+C_{loss}-\sum_{i \in \mathcal{I}} C_{g,i}
\end{aligned}
\end{equation}
where 
\begin{align*}
    C_{loss}=&\sum_{t\in\mathcal{T}}\sum_{(i,j)\in\mathcal{E}}r_{ij}\ell_{ij}(t)\pi(t),\\
    C_{g,i}=& \sum_{t \in \mathcal{T}} p_{g,i}(t)\lambda_{p,i}(t)\Delta t= \sum_{t \in \mathcal{T}} p_{n_i}(t)\lambda_{p,i}(t)\Delta t,
\end{align*}
where $\pi(t)$ turns the power loss into a monetary term. Vector $\mathbf{x_d}=\{\mathbf{p,q,P,Q,\ell,v}\}$ summarizes the decision variables, $\mathbf{p}=\{\mathbf{p_n(t)},n\in\mathcal{N},t\in\mathcal{T}\}$,
$\mathbf{q}=\{\mathbf{q_n(t)},n\in\mathcal{N},t\in\mathcal{T}\}$,
$\mathbf{P}=\{\mathbf{P_{nj}(t)}, (n,j)\in\mathcal{E},t\in\mathcal{T}\}$,
$\mathbf{Q}=\{\mathbf{Q_{nj}(t)}, (n,j)\in\mathcal{E},t\in\mathcal{T}\}$,
$\mathbf{\ell}=\{\mathbf{\ell_{nj}(t)}, (n,j)\in\mathcal{E},t\in\mathcal{T}\}$,
$\mathbf{v}=\{\mathbf{v_{n}(t)}, n\in\mathcal{N},t\in\mathcal{T}\}$.

\subsection{Distributed Coordination Mechanism}\label{sec:form}
According to the above models, we can find that the charging stations and the distribution network interact with each other through equation (\ref{equ:pi}), which acts as a bridge connecting the charging stations and the distributed network.
In the following, we design a coordination mechanism by modifying the objective functions (\ref{equ:objci}) and (\ref{equ:objdso}) to take into account the connecting equation (\ref{equ:pi}). The coordination mechanism will run in an iterative process. In the $(k+1)$-th iteration:

The objective function of charging station $i$ is modified as
\begin{equation}\label{equ:objci1}
\begin{aligned}
C_i(\mathbf{x_i}) =& C_{b,i} +C_{ev,i}+C_{g,i}+\sum_{t\in\mathcal{T}}\frac{\rho}{2}(p_{g,i}(t)-p_{n_i}^{k}(t))^2\\
=& C_{b,i} +C_{ev,i}+\sum_{t\in\mathcal{T}}\left(p_{g,i}(t)\lambda_{p,i}^k(t)\right)\Delta t\\
&+\sum_{t\in\mathcal{T}}\frac{\rho}{2}(p_{g,i}(t)-p_{n_i}^{k}(t))^2,
\end{aligned}
\end{equation}
subject to 
\begin{equation*}
(\ref{equ:evcons}),(\ref{equ:pbdch})-(\ref{equ:Ebatt}),(\ref{equ:pgb}),(\ref{equ:pBala}),
\end{equation*}
where $\{p_{n_i}^k(t),\forall t\}$ and $\{\lambda_{p,i}^k(t),\forall t\}$ are the given desired energy trading schedule and energy trading prices announced by the DSO, respectively. The last quadratic term in the objective function represents the gap between the charging station $i$'s actual energy trading profile $\{p_{g,i}(t),\forall t\}$ and the DSO's schedule. The charging station $i$ needs to meet DSO's schedule as much as possible. Denote the charging station's optimal energy trading profile as $\{p_{g,i}^{k+1}(t),\forall t\}$.


Meanwhile, for DSO, its objective function is modified as
\begin{align}\label{equ:objdso1}
C_{dso}(\mathbf{x_d}) =& C_{bus1} +C_{loss}-\sum_{i \in \mathcal{I}} C_{g,i} \nonumber\\
& +\sum_{i \in \mathcal{I}}\sum_{t\in\mathcal{T}}\frac{\rho}{2}(p_{g,i}^{k+1}(t)-p_{n_i}(t))^2 \nonumber\\
=& C_{bus1}+C_{loss}-\sum_{i \in \mathcal{I}}\sum_{t\in\mathcal{T}}\left(p_{n_i}(t)\lambda_{p,i}^k(t)\right)\Delta t \nonumber\\
&+\sum_{i \in \mathcal{I}}\sum_{t\in\mathcal{T}}\frac{\rho}{2}(p_{g,i}^{k+1}(t)-p_{n_i}(t))^2,
\end{align}
subject to
\begin{equation*}
(\ref{eq:ACpowerflow})-(\ref{equ:pi}),
\end{equation*}
where $\{p_{g,i}^{k+1}(t),\forall i,\forall t\}$ are the desired energy trading profiles submitted by the charging stations. The DSO will decide on the energy trading schedule $\{p_{n_i}(t),\forall i,\forall t\}$ to meet the needs of charging stations as much as possible, which is represented in the last quadratic term of the modified objective function \eqref{equ:objdso1}. After determining the energy trading schedule $\{p_{n_i}^{k+1}(t),\forall i,\forall t\}$ by solving the above optimization problem, the DSO will then update the energy trading prices $\{\lambda_{p,i}^{k+1}(t),\forall i,\forall t\}$. Consider that the obtained $p_{n_i}^{k+1}(t)$ may not equal to the charging station's needs $p_{g,i}^{k+1}(t)$, the DSO will adjust the prices to reduce this power imbalance by
\begin{align}
\lambda^{k+1}_{p,i}(t)&=\lambda^{k}_{p,i}(t)+\rho\left(p^{k+1}_{g,i}(t)-p^{k+1}_{n_i}(t)\right).\label{equ:step3}
\end{align}
When the power needed by the charging station $p_{g,i}^{k+1}(t)$ is less than the DSO's desired schedule $p_{n_i}^{k+1}(t)$, i.e., $p_{g,i}^{k+1}(t)-p_{n_i}^{k+1}(t)<0$, the DSO will lower the energy trading price ($\lambda_{p,i}^{k+1}(t)<\lambda_{p,i}^k(t)$) to incentivize the charging station to buy more electricity, and vice versa.

The interaction between the charging stations and the DSO happens iteratively, and the mechanism stops when the energy trading price becomes almost stable, i.e., the price difference between two successive iterations is less than the tolerance $\delta$, i.e.,
\begin{equation}\label{equ:r}
  r=\|\mathbf{\lambda}^{k+1}_p-\mathbf{\lambda}^{k}_p\|\leq\delta.
\end{equation}

A completed description of the proposed coordination mechanism is shown in Algorithm \ref{algo:dc}. The data exchanged between the charging stations and the DSO consists solely of the energy trading schedule/profile and the trading prices. Therefore, the private data of individual charging stations and the network parameters owned by the DSO are well protected.

\begin{algorithm}[htbp]
\caption{Distributed Coordination Mechanism}\label{algo:dc}
\begin{algorithmic}[1]
\STATE{Set iteration index $k=0$, convergence error tolerance $\delta>0$, penalty parameter $\rho>0$.}
\STATE{DSO initializes the energy trading price $\lambda^k_{p,i}=0$, and desired traded energy $p^k_{n_i}=0,i\in\mathcal{I}$  for charging stations.}
\REPEAT
\FOR{Each charging station $i\in\mathcal{I}$}
\STATE{CSO updates $p_{g,i}^{k+1}$ according to the modified objective function (\ref{equ:objci1}), and sends them to DSO.}
\ENDFOR
\STATE DSO then updates $p_{n_i}^{k+1},\forall i$ according to the modified objective function (\ref{equ:objdso1}), and the trading prices $\lambda^{k+1}_{p,i},\forall i$ via (\ref{equ:step3}), and broadcasts them to each CSO.
\STATE {Set $k=k+1$}
\UNTIL {convergence stopping criterion (\ref{equ:r}) is satisfied.}
\end{algorithmic}
\end{algorithm}

Still, we have three concerns about the above proposed coordination mechanism:
1) The quadratic equality constraint (\ref{equ:pfnonconvex}) is nonconvex, making the DSO's problem challenging to solve.
2) Is there any convergence guarantee for the proposed mechanism? 3) When it converges, will the equilibrium be close to or the same as the centralized optimum?

For the first issue, convex relaxation is performed to turn (\ref{equ:pfnonconvex}) into a second-order conic inequality constraint, i.e.,
\begin{equation}\label{equ:socp}
\ell_{ij}(t)\geq\frac{P_{ij}(t)^2+Q_{ij}(t)^2}{v_i(t)}.
\end{equation}
The exactness of relaxation has been verified for the radial distribution network \cite{low2013branch} when the power injection at each bus is not too large, and the bus voltage maintains around the nominal value.

For the rest two issues, we have
\begin{proposition}\label{prop-2}
The proposed distributed coordination mechanism is guaranteed to converge to the centralized optimum.
\end{proposition}

The proof of Proposition \ref{prop-2} can be found in Appendix \ref{appendix-B}.


\section{Simulation Results and Discussion}\label{sec:result}
\begin{figure}[!htbp]
  \centering
  \includegraphics[width=0.4\textwidth]{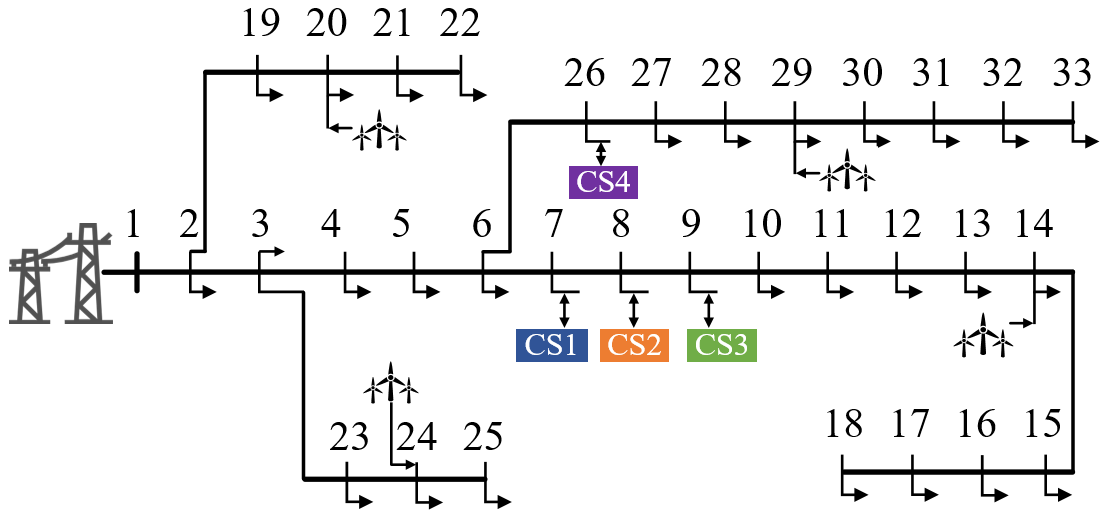}\\
  \caption{Modified IEEE 33-bus test system.}\label{fig:ieee33}
\end{figure}

In this section, we evaluate the performance of the proposed two-stage scheme, including the aggregate EV power flexibility evaluation and the distributed coordination mechanism using a IEEE 33-bus test system \cite{ieee33} with four charging stations: CS1, CS2, CS3, and CS4, as shown in Fig. \ref{fig:ieee33}.

\subsection{System Setup}
We consider a system with four charging stations. They are all equipped with PV panels and batteries, but their capacities are different.
The related batteries parameters are $E_{b,1}/E_{b,2}/E_{b,3}/E_{b,4}=100/150/200/200\text{ kWh}, p^{c,\max}_{b,1}=p^{d,\max}_{b,1}=30\text{ kW},
p^{c,\max}_{b,2}=p^{d,\max}_{b,2}=45\text{ kW},
p^{c,\max}_{b,3}=p^{d,\max}_{b,3}=60\text{ kW},
p^{c,\max}_{b,4}=p^{d,\max}_{b,4}=60\text{ kW},
\eta_d=\eta_c=0.95,  c_{b,1}/c_{b,2}/c_{b,3}/c_{b,4}=0.1.$
The utility grid's hourly buy electricity price profile $\lambda^b_{g}$ is obtained from the PJM electricity market~\cite{pjm}.
The sale electricity price to the utility grid $\lambda^s_{g}$ is set to be a constant value 0.01 \$/kWh.
The maximum allowable traded power of the charging station is set as $p_{g,i}^{b,\max}=p_{g,i}^{s,\max}=300\text{ kW},\forall i$.
In general, the number of EVs charging at a charging station reflects how busy the station is. If a charging station has more EV charging tasks, its charging pattern is more severe, and vice versa. 
We consider four charging stations with different busy levels: mild, moderate, and severe. The number of chargers in each station $N_{chg,i}$ is 20. The daily number of EVs charging at station $i$ is $|\mathcal{V}_i|$. A bigger ratio $\frac{|\mathcal{V}_i|}{N_{chg,i}}$ means a severer charging pattern.
For CS1, there are $|\mathcal{V}_1|=30$ EVs during 6:00-22:00, i.e., $\frac{|\mathcal{V}_1|}{N_{chg,1}}=1.5$, representing a moderate charging pattern. The CS2 has a mild charging pattern, in which there are $|\mathcal{V}_2|=20$ EVs during 6:00-22:00, i.e., $\frac{|\mathcal{V}_2|}{N_{chg,2}}=1$. The CS3 also has a moderate charging pattern with $|\mathcal{V}_3|=30$ ($\frac{|\mathcal{V}_3|}{N_{chg,3}}=1.5$), but with 15 EVs during 4:00-14:00 and 15 EVs during 14:00-23:00. The CS4 has a severe charging pattern, $|\mathcal{V}_4|=40$ ($\frac{|\mathcal{V}_4|}{N_{chg,4}}=2$), and there are 10 EVs during 2:00-8:00, 10 EVs during 8:00-14:00, 10 EVs during 14:00-20:00, and 10 EVs during 20:00-23:00.
Other EV charging related parameters are $E_{v}=40~\text{kWh}, soc^{\min}_v=0.1, soc^{\max}_v=0.9, soc^{ini}_v=0.2, soc^{req}_v=0.5, c_{cs,i}=0.1$.
For the chargers, we have $p_{chg}=6.6~\text{kW}$.
We set the upper and
lower limits of voltage magnitude in each bus as 1.06 p.u. and 0.94 p.u., respectively.

\subsection{Aggregate EV Power Flexibility}
\begin{figure}[!htbp]
  \centering  \includegraphics[width=0.4\textwidth]{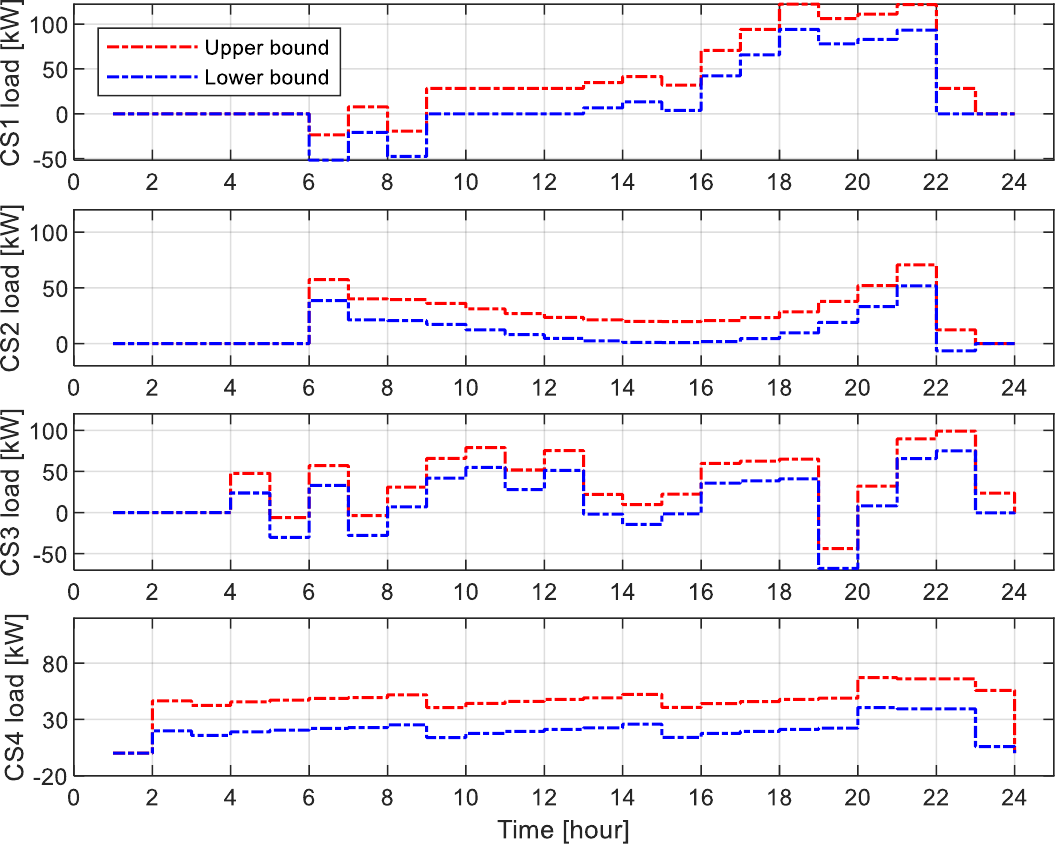}\\
  \caption{Maximal aggregate EV power flexibility region.}\label{fig:mf}
\end{figure}

\begin{figure}[!htbp]
  \centering  \includegraphics[width=0.4\textwidth]{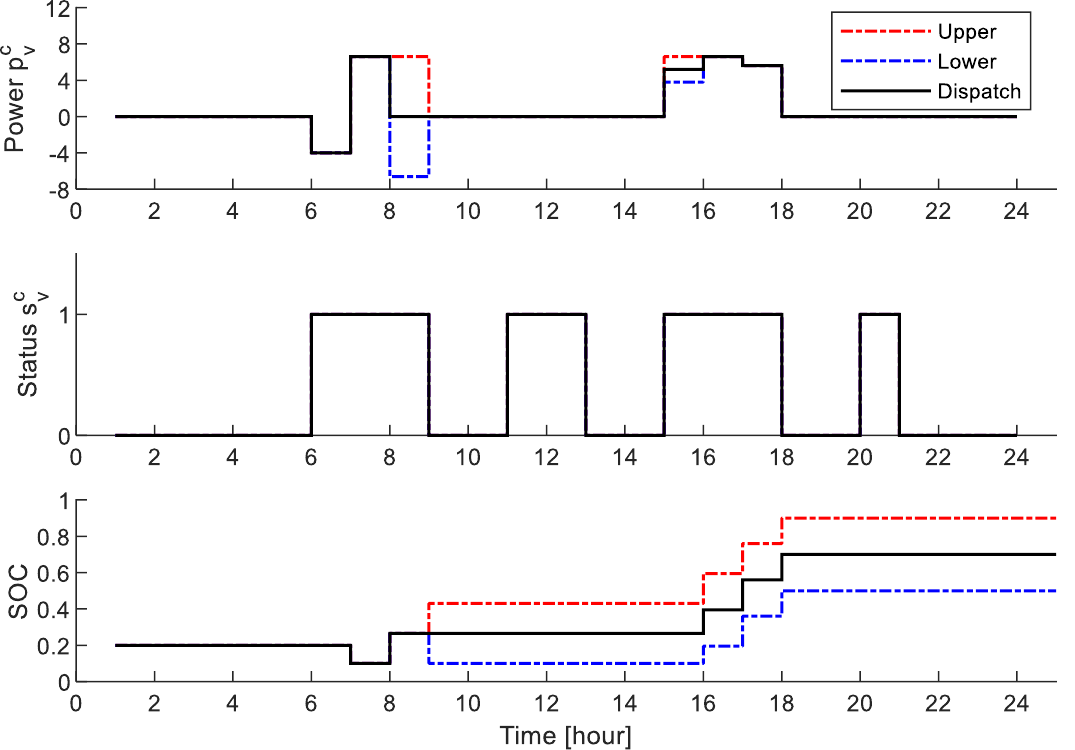}\\
  \caption{Realized dispatch of an EV.}\label{fig:adispatch}
\end{figure}

\begin{figure}[!htbp]
  \centering  \includegraphics[width=0.4\textwidth]{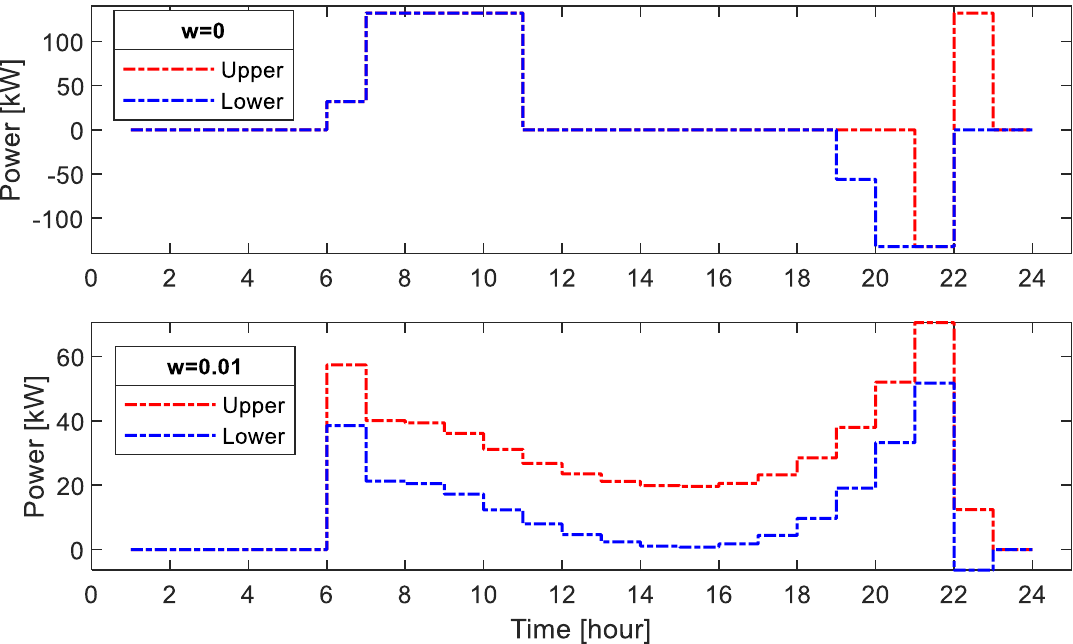}\\
  \caption{Impact of $w$ on power flexibility region.}\label{fig:wlbub}
\end{figure}

Fig. \ref{fig:mf} shows the obtained aggregate EV power flexibility regions of the four charging stations based on the first stage problem.
The area between the upper and lower power trajectories is the aggregate power flexibility that is available to dispatch by charging stations in the second stage.
Within this area, any power trajectory is dispatchable and can satisfy the EVs' charging requirements. Here, we randomly pick up one EV power trajectory in CS1 and recover its dispatch in Fig.~\ref{fig:adispatch}. The charging power, charging status, and SOC all satisfy the constraints, which validates Proposition \ref{prop-1}.
Additionally, Fig.~\ref{fig:wlbub} shows the impact of the quadratic term in (\ref{equ:drobj}) on the flexibility region. In the case of $w=0$, i.e., without the quadratic term, the power flexibility region becomes more concentrated in some periods. On the contrary, with the help of quadratic term (e.g., $w=0.01$), the power flexibility region is more evenly distributed. This provides more available dispatchable opportunities for charging station.

\subsection{Exactness of the Second-Order Conic Relaxation and Convergence of the Proposed Coordination Mechanism}
\begin{figure}[!htbp]
  \centering
  \includegraphics[width=0.4\textwidth]{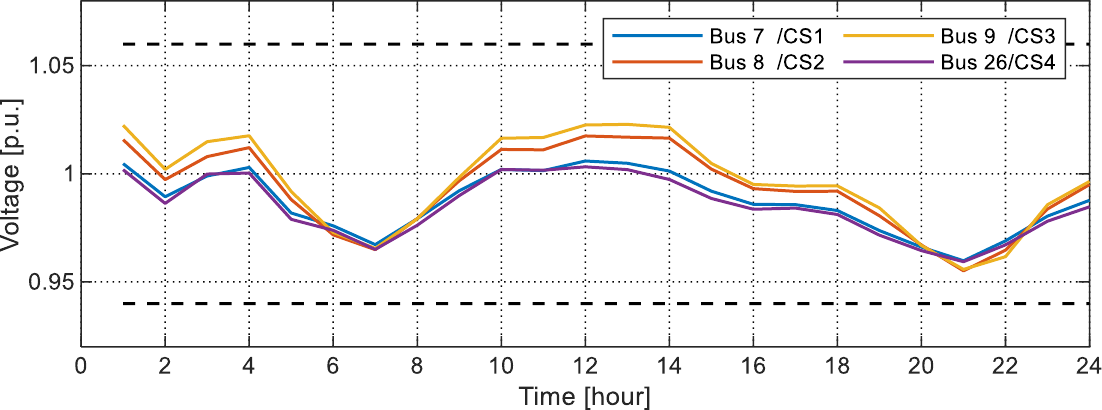}\\
  \caption{Bus voltage at each charging station.}\label{fig:volt}
\end{figure}

\begin{figure}[!htbp]
  \centering
  \includegraphics[width=0.40\textwidth]{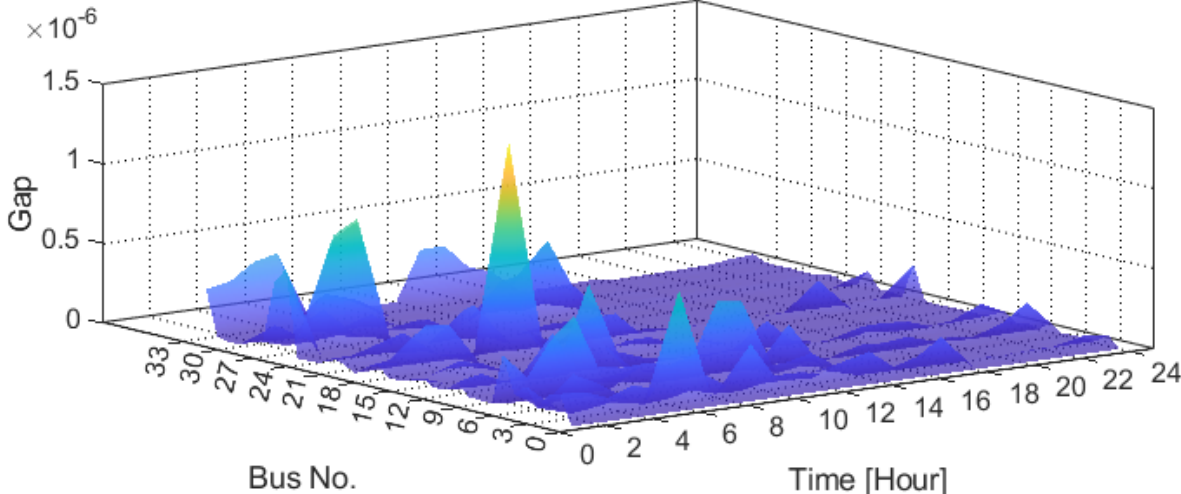}\\
  \caption{The gap at each bus under different periods.}\label{fig:vgap}
\end{figure}

With the aggregate EV power flexibility obtained in the first stage, we then run the coordination mechanism proposed in the second stage to settle the energy transactions.
At the equilibrium point, the voltages of buses with charging stations are shown in Fig. \ref{fig:volt}.
We can see that all the voltages are within the allowable range. In addition, to check the exactness of the second-order conic relaxation, we also illustrate the difference between the left hand side $\ell_{i,j}(t)$ and the right hand side $\frac{P_{ij}(t)^2+Q_{ij}(t)^2}{v_i(t)}$ of the inequality (\ref{equ:socp}).
As shown in Fig. \ref{fig:vgap}, all gap values are less than $10^{-6}$, which is very small. That is, equality almost holds for the inequality constraint (\ref{equ:socp}), so the second-order conic relaxation is exact.

\begin{figure}[!htbp]
  \centering
    \includegraphics[width=0.48\textwidth]{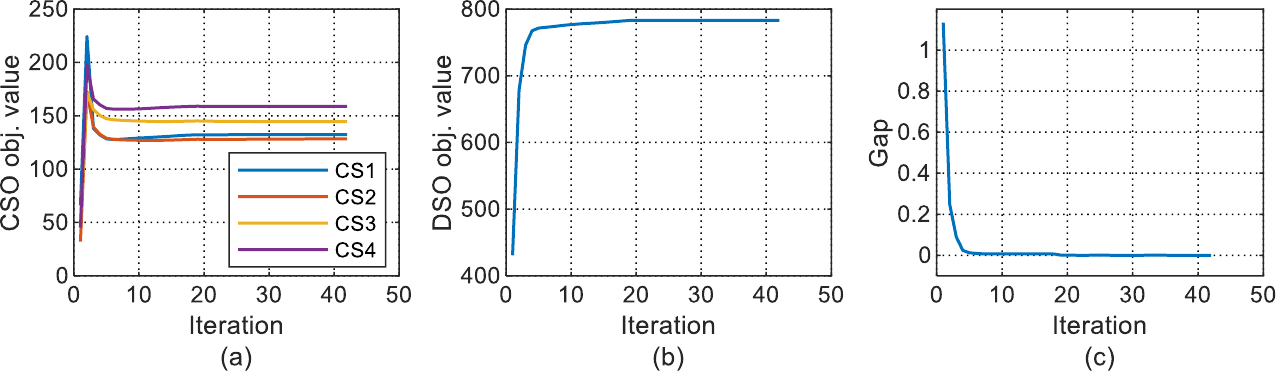} \\
  \caption{Convergence illustration. (a) Charging stations' objective function values. (b) DSO's objective function value. (c) Error/gap $r$.}\label{fig:converge}
\end{figure}

Fig. \ref{fig:converge} shows the convergence of the proposed distributed coordination mechanism.
It can be found that the number of iterations to reach convergence is about 42, which indicates that the convergence speed is fast and acceptable.

\subsection{Performance Evaluation}
To demonstrate the advantages of the proposed mechanism, the conventional method (\ref{equ:cgiBase}) is used as a baseline. In the baseline, each charging station operates individually to calculate the input/output power from/to the grid. Then the DSO manages the power flow of the distribution network by solving the following problem:
\begin{equation}
\begin{aligned}
    \min~ & C^{base}_{dso}(\mathbf{x_{d}})
    =C_{bus1}+C_{loss}-\sum_{i \in \mathcal{I}} C^{base}_{g,i}\\
    \text{s.t.}~ & (\ref{equ:pinpf})-(\ref{equ:pfvj}),(\ref{equ:socp}),(\ref{equ:pfpqub}),(\ref{equ:pflvub}),\\
    &p_{n_i}(t) = p^b_{g,i}(t)-p^s_{g,i}(t).
\end{aligned}
\end{equation}
\subsubsection{Cost Comparison}
TABLE \ref{tab:compare} summarizes the costs under the two different mechanisms.
The total cost of each charging station $C_i$ comprises the energy trading cost $C_{g,i}$, battery operation cost $C_{b,i}$, and EV dissatisfaction cost $C_{ev,i}$.
The proposed mechanism (212 USD) performs better than the baseline (293.36 USD) with a significant total cost reduction (27.73\%).
In particular, under the proposed mechanism, the energy trading cost $C_{g,i}$ for charging stations (CS2, CS3, and CS4) are negative, which means that they can earn revenue from selling energy to other charging stations or the utility grid instead of paying more on buying energy from the utility grid under the baseline. For the DSO, the proposed mechanism can also achieve a lower total cost (517.86 USD) than the baseline (547.89 USD).
Specifically, compared with the baseline, the proposed mechanism can reduce the amount of energy bought from the utility grid.
In addition, the power loss cost of the proposed method is also lower than that of the baseline. This reduction in power loss benefits the transmission line.
Finally, as seen in the table, the proposed mechanism (729.87 USD) outperforms the baseline (841.24 USD) in terms of the total cost of the overall system, with a great decrease (13.24\%).
In conclusion, the proposed mechanism can not only benefit the charging stations, but also the distribution network by reducing power losses and costs, resulting in a win-win situation.

\begin{table*}[!hbtp]
  \centering
  \caption{Cost comparison between baseline and proposed mechanism (Unit: USD).}\label{tab:compare}
    \begin{tabular}{c|cccccc|cccc|cc}
    \hline\hline
          &       & \multicolumn{5}{c|}{Charging station} & \multicolumn{4}{c|}{Distribution network} & \multirow{2}[2]{*}{Total cost} & \multirow{2}[2]{*}{Reduction} \\
          &       & $C_{g,i}$ & $C_{b,i}$ & $C_{ev,i}$ & $C_{i}$   & $\sum_{i\in\mathcal{I}} C_{i}$ & $C_{bus1}$ & $C_{loss}$ & $C_{g,i}$ & $C_{dso}$ &       &  \\
    \hline
    \multirow{4}[1]{*}{Baseline} & CS1   & 71.71 & 12.65 & 21.99 & 106.35 & \multirow{4}[1]{*}{293.36} & \multirow{4}[1]{*}{610.39} & \multirow{4}[1]{*}{111.68} & -71.71 & \multirow{4}[1]{*}{547.89 } & \multirow{4}[1]{*}{841.24 } & \multirow{4}[1]{*}{-} \\
          & CS2   & 22.78 & 0.04  & 8.80  & 31.62 &       &       &       & -22.78 &       &       &  \\
          & CS3   & 36.82 & 11.66 & 26.33 & 74.81 &       &       &       & -36.82 &       &       &  \\
          & CS4   & 42.88 & 2.97  & 34.73 & 80.57 &       &       &       & -42.88 &       &       &  \\
    \hline
    \multirow{4}[1]{*}{Proposed} & CS1   & 53.20 & 19.02 & 31.06 & 103.28 & \multirow{4}[1]{*}{212.00} & \multirow{4}[1]{*}{402.53} & \multirow{4}[1]{*}{87.76} & -53.20 & \multirow{4}[1]{*}{517.86 } & \multirow{4}[1]{*}{729.87 } & \multirow{4}[1]{*}{13.24\%} \\
          & CS2   & -38.87 & 28.53 & 20.26 & 9.92  &       &       &       & 38.87 &       &       &  \\
          & CS3   & -29.52 & 38.04 & 27.73 & 36.26 &       &       &       & 29.52 &       &       &  \\
          & CS4   & -12.39 & 38.04 & 36.89 & 62.54 &       &       &       & 12.39 &       &       &  \\
    \hline\hline
    \end{tabular}%
\end{table*}%

\subsubsection{Impact on Power Loss}
\begin{figure}[!htbp]
  \centering
  \includegraphics[width=0.4\textwidth]{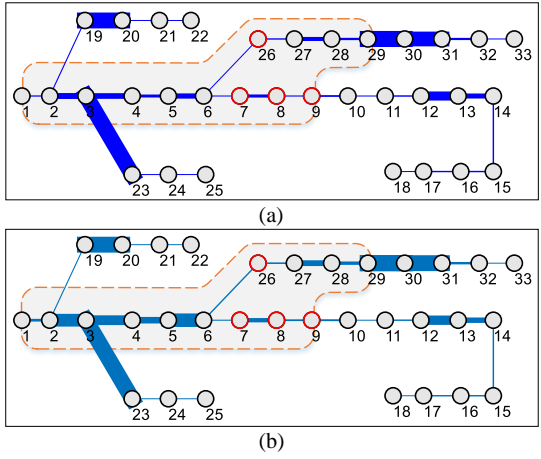}\\   \caption{Visualization of the power loss of each line under two mechanisms. (a) Proposed. (b) Baseline.}\label{fig:loss}
\end{figure}

To understand the impact of energy trading on the power flow of the distribution network, Fig.~\ref{fig:loss} visualizes the power loss on each line by using the line width to represent the amount of power loss. The thicker the line, the greater the power loss. In the circled area covering the charging stations and bus 1, the proposed method (Fig.~\ref{fig:loss} (a)) generates much thinner lines than the baseline (Fig.~\ref{fig:loss} (b)). This indicates less power loss in these lines, which is beneficial to prolong the service life of transmission lines. It can be concluded that the proposed coordination mechanism can 
promote local transactions and reduce the power loss of the distribution network.

\subsubsection{Energy Trading Price}
\begin{figure}[!htbp]
  \centering
  \includegraphics[width=0.4\textwidth]{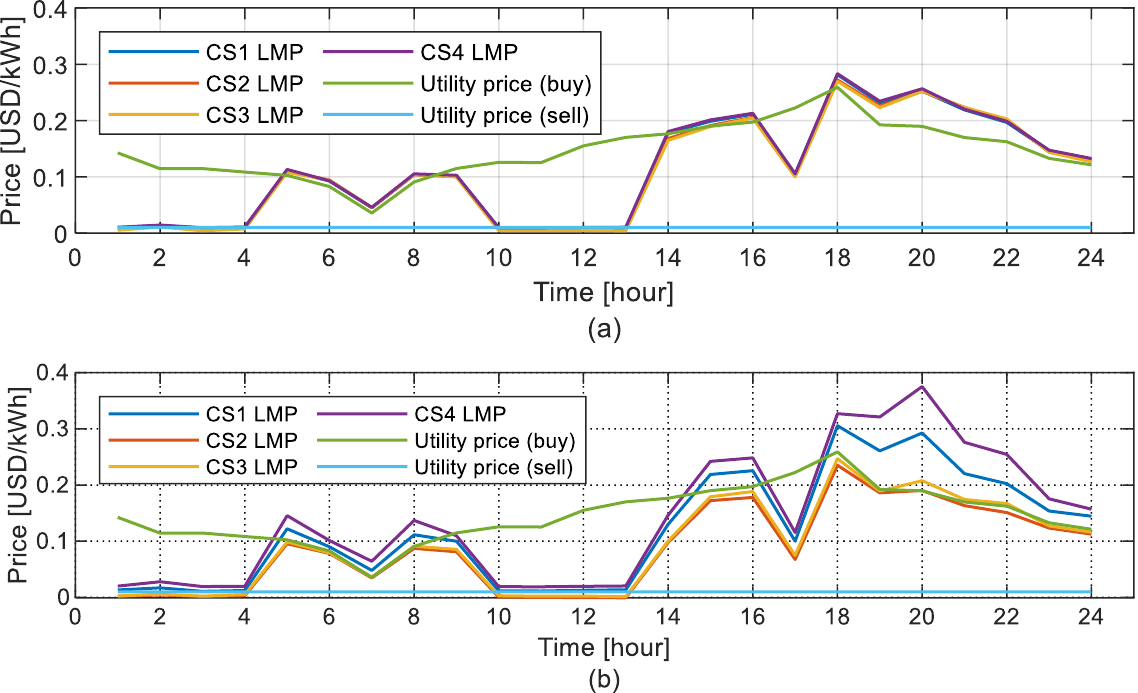}\\
  \caption{The energy trading prices. (a) CSs are located close to each other. (b) CSs are far away from each other.}\label{fig:lmp}
\end{figure}

Since the energy trading cost accounts for the majority of the overall cost of the charging stations in both the proposed mechanism and the baseline, it would be interesting to analyze the energy trading prices underlying the outcomes.
As mentioned before, under the proposed mechanism, the energy trading price is determined by the dual variable of \eqref{equ:pi}, which well reflects the value of electricity at different buses, accounting for the load and generation levels, and the physical limits of the transmission lines, etc.
Fig. \ref{fig:lmp} (a) shows the energy trading price profiles of four charging stations. First, the trading prices of the four charging stations follow the same trend. This is because they are located close to each other. As a counter example, a case that the four charging stations are located far from each other (located at bus 7, 22, 25, and 33) is also simulated. Fig.~\ref{fig:lmp} (b) shows their trading prices. As seen, the trading prices are different and divergent. 

\subsubsection{Internal Power Distribution of Each Charging Station}
\begin{figure}[!htbp]
  \centering
    \includegraphics[width=0.4\textwidth]{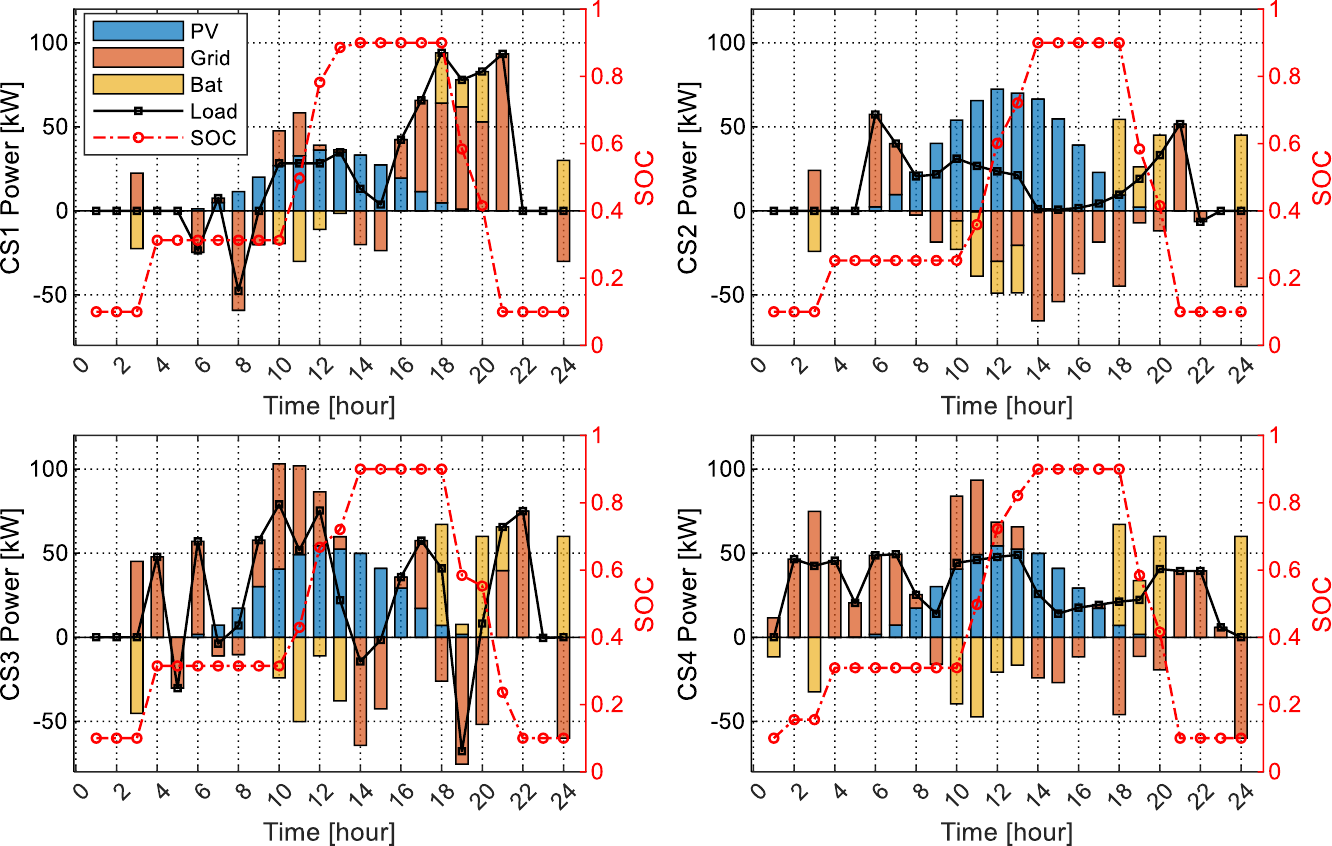} \\
  \caption{Proposed: Power distribution inside each charging station.}\label{fig:power}
\end{figure}

\begin{figure}[!htbp]
  \centering
  \includegraphics[width=0.4\textwidth]{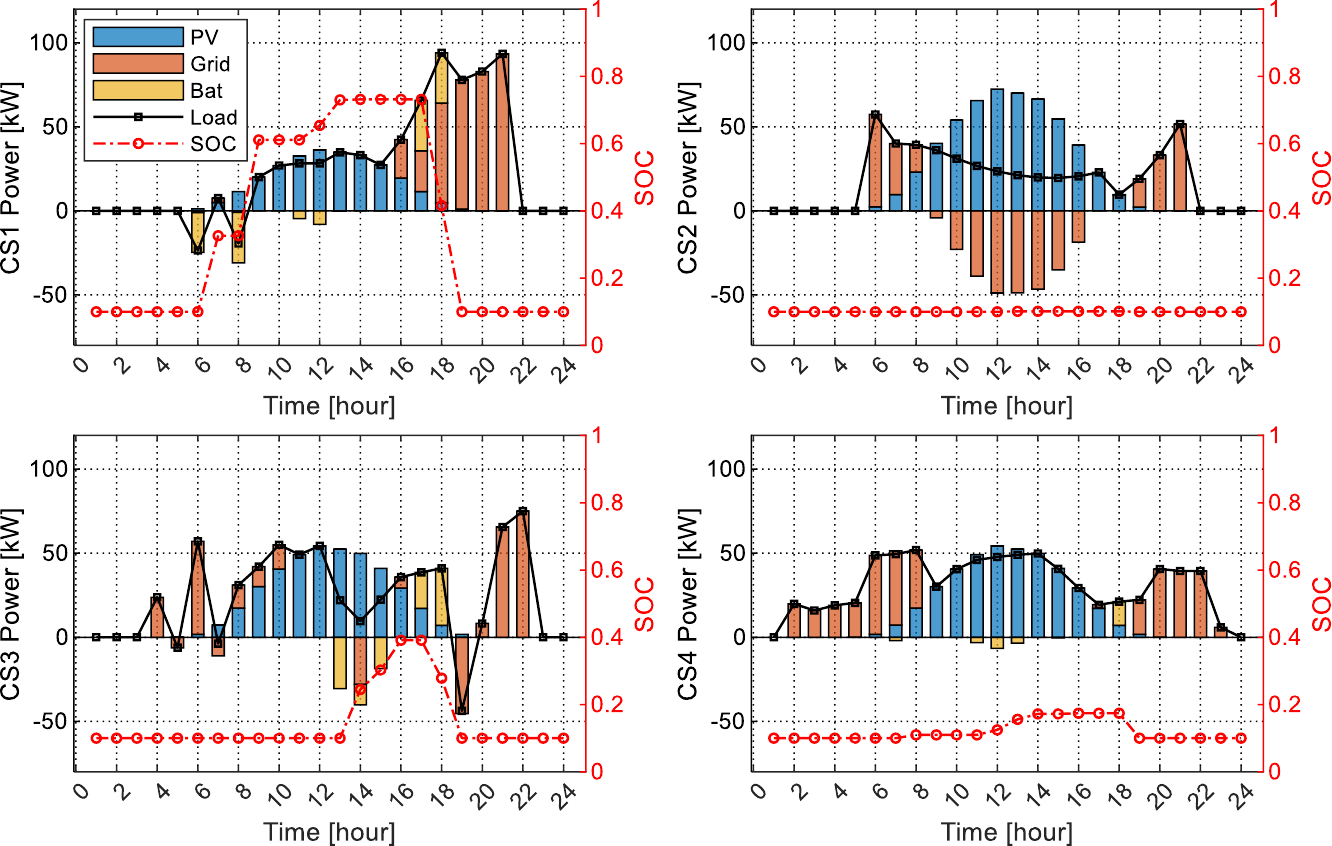} \\
  \caption{Baseline: Power distribution inside the four charging stations.}\label{fig:basepower}
\end{figure}

Fig. \ref{fig:power} shows the charging station's internal power distribution among the grid, battery energy storage, and PV, as well as the SOC evolution of battery energy storage under the proposed mechanism.
For comparison, Fig. \ref{fig:basepower} shows the power distribution under the baseline, which presents distinct features from Fig. \ref{fig:power}.
A major difference is that: according to the SOC level, the battery storage was not fully utilized in the baseline, particularly for CS2-CS4; while under the proposed mechanism, the battery SOCs of the four charging stations vary in the same trend and experience a complete charging and discharging cycle.
Thus, it can be deduced that with the proposed properly designed coordination mechanism and energy trading price, the utilization of energy storage can be greatly promoted. 

\subsubsection{Scalability}
\begin{figure}[!htbp]
  \centering
  \includegraphics[width=0.4\textwidth]{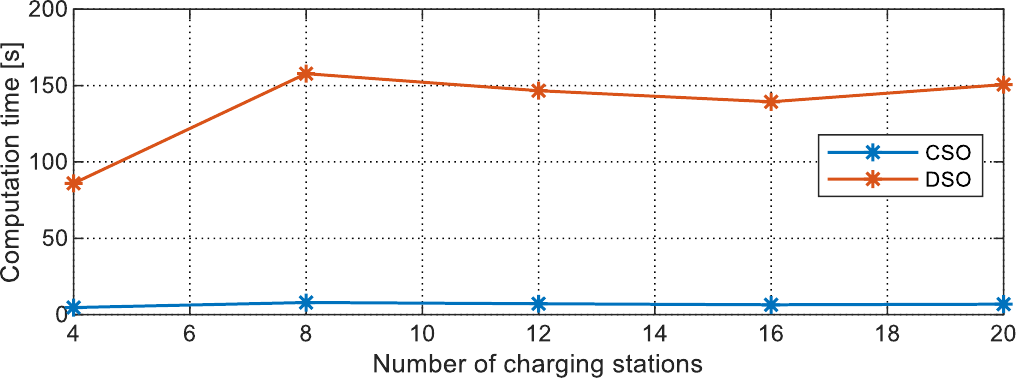} \\
  \caption{Scalability: Computational time vs. number of charging stations.}\label{fig:scala}
\end{figure}

To show the scalability of the proposed mechanism, we compare the computational time under different number of charging stations: 4, 8, 12, 16, and 20. 
As seen in Fig.~\ref{fig:scala}, when the number of charging stations increases, the total computational time of a CSO and DSO remains about 7s and 150s, respectively. This is because the calculations of charging stations can be run in parallel. Therefore, the proposed mechanism is not sensitive to the number of charging stations.

\section{Conclusion}\label{sec:conclu}
This paper proposes a two-stage scheme to coordinate the EV charging stations. In contrast to the existing approaches that assume the charging station has complete information about the EVs, we use an aggregate EV power flexibility region instead. This can protect EV owners' privacy and be much easier for the charging station to dispatch.
In the second stage, a novel distributed coordination mechanism is proposed. We prove that the proposed mechanism can converge to the centralized optimum.
Simulations demonstrate the effectiveness of the proposed scheme and have the following findings:

(1) Compared with the traditional method using a uniform price, the proposed coordination mechanism can achieve a significant total cost reduction of 13.24\%.

(2) The energy sharing of charging stations affects the local power flow of the distribution network and results in a lower power loss.

(3) Energy trading enhances the utilization rate of battery storage deployed in the charging stations.

\bibliographystyle{IEEEtran}
\bibliography{PaperRef}

\appendix
\makeatletter
\@addtoreset{equation}{section}
\@addtoreset{theorem}{section}
\makeatother
\setcounter{equation}{0}  
\renewcommand{\theequation}{A.\arabic{equation}}
\renewcommand{\thetheorem}{A.\arabic{theorem}}

\subsection{Proof of Proposition \ref{prop-1}}
\label{appendix-A}
For each time slot $t \in \mathcal{T}$, since $p_{d,i}(t) \in [\check p_{d,i}^*(t), \hat p_{d,i}^*(t)]$, we can define an auxiliary coefficient:
\begin{align}
    \alpha(t):= \frac{\hat p_{d,i}^*(t)-p_{d,i}(t)}{\hat p_{d,i}^*(t)-\check p_{d,i}^*(t)} \in [0,1]
\end{align}
so that $p_{d,i}(t)=\alpha(t) \check{p}^*_{d,i}(t)+(1-\alpha(t))\hat{p}^*_{d,i}(t)$. Then, we can construct a feasible EV dispatch strategy by letting
\begin{subequations}
\begin{align}
    p^c_v(t)&=\alpha(t)\check{p}^{c*}_{v}(t)+(1-\alpha(t))\hat{p}^{c*}_{v}(t),\\
    s^c_v(t)&=\alpha(t)\check{s}^{c*}_{v}(t)+(1-\alpha(t))\hat{s}^{c*}_{v}(t),\\
    soc_v(t)&=\alpha(t)\check{soc}^{c*}_{v}(t)+(1-\alpha(t))\hat{soc}^{c*}_{v}(t).
\end{align}
\end{subequations}
for all time slots $t \in \mathcal{T}$.

We prove that it is a feasible EV dispatch strategy as follows. First, due to \eqref{equ:joint} we have $s_v^c=\check s_v^c=\hat s_v^c$, thus, constraints \eqref{equ:ubscsd} and \eqref{equ:ubnchg1} hold for $s_v^c(t)$. Furthermore,
\begin{align}
    p_{d,i}(t)=~ & \alpha(t)\check p_{d,i}^*(t) + (1-\alpha(t)) \hat p_{d,i}^*(t) \nonumber\\
    =~ &  \alpha(t) \sum_{v \in \mathcal{V}_i} \check p_v^{c*}(t) + (1-\alpha(t)) \sum_{v \in \mathcal{V}_i} \hat p_v^{c*}(t) \nonumber\\
    = ~ & \sum_{v \in \mathcal{V}_i} \left[\alpha(t) \check p_v^{c*}(t)+(1-\alpha(t)) \hat p_v^{c*}(t)\right] \nonumber\\
    = ~ & \sum_{v \in \mathcal{V}_i} p_v^c(t)
\end{align}
Hence, constraint \eqref{equ:ubpdi} holds for $p_{d,i}(t)$ and $p_v^c(t),\forall v$. Similarly, we can prove that constraints \eqref{equ:ubpd}, \eqref{equ:ubsntatd}-\eqref{equ:ubsocranges} are met. Therefore, we have constructed a feasible EV dispatch strategy, which completes the proof. \hfill$\blacksquare$

\setcounter{equation}{0}  
\renewcommand{\theequation}{B.\arabic{equation}}
\renewcommand{\thetheorem}{B.\arabic{theorem}}

\subsection{Proof of Proposition \ref{prop-2}}
\label{appendix-B}
For objective function (\ref{equ:objci1}), if we add a constant term $-p_{n_i}^k(t)\lambda_{p,i}^k(t)\Delta t$ to it, this won't affect the optimal solution of the problem. Then we can reorganize the objective function as follows
\begin{equation}\label{equ:objci2}
\begin{aligned}
C_i(\mathbf{x_i})
=& C_{b,i} +C_{ev,i}+\sum_{t\in\mathcal{T}}\left[p_{g,i}(t)-p_{n_i}^k(t)\right]\lambda_{p,i}^k(t)\Delta t\\
&+\sum_{t\in\mathcal{T}}\frac{\rho}{2}(p_{g,i}(t)-p_{n_i}^k(t))^2.
\end{aligned}
\end{equation}

Similarly, we add a constant term $p_{g,i}^{k+1}(t)\lambda_{p,i}^k(t)\Delta t$ to the objective function (\ref{equ:objdso1}), and reorganize it into
\begin{equation}\label{equ:objdso2}
\begin{aligned}
C_{dso}(\mathbf{x_d})
=& C_{bus1}+C_{loss}+\sum_{t\in\mathcal{T}}\left[p_{g,i}^{k+1}(t)-p_{n_i}(t)\right]\lambda_{p,i}^k(t)\Delta t\\
&+\sum_{t\in\mathcal{T}}\frac{\rho}{2}(p_{g,i}^{k+1}(t)-p_{n_i}(t))^2.
\end{aligned}
\end{equation}

After the equivalent transformations, we can find that our proposed mechanism turns into the classical ADMM framework. It has been proven that the distributed ADMM framework has good convergence performance and can converge to a centralized optimum if the problem is convex. Recall that we have carried out convex relaxation (\ref{equ:socp}) in dealing with the first issue. Therefore, we can easily prove Proposition \ref{prop-2} following a similar procedure as the ADMM framework.  \hfill$\blacksquare$

\end{document}